\documentstyle[12pt,aaspp4,epsfig]{article}

\begin{document}

\title{Bayesian photometric redshift estimation}
\author{Narciso Ben\'\i tez}
\affil{Astronomy Department, UC Berkeley, 601 Campbell Hall, 
Berkeley, CA}
\authoremail{benitezn@wibble.berkeley.edu}

\begin{abstract}

Photometric redshift estimation is becoming an increasingly important 
technique, although the currently existing methods present several 
shortcomings which hinder their application. Here it is shown that 
most of those drawbacks are efficiently eliminated when Bayesian 
probability is consistently applied to this problem. The use of prior 
probabilities and Bayesian marginalization allows the inclusion of 
valuable information, e.g. the redshift distributions or the galaxy type 
mix, which is often ignored by other methods. It is possible to quantify the 
accuracy of the redshift estimation in a way with no equivalents in 
other statistical approaches; this property permits the selection of 
galaxy samples for which the redshift estimation is extremely reliable. 
In those cases when the {\it a priori} information is insufficient, it 
is shown how to `calibrate' the prior distributions, using even the data 
under consideration. 

There is an excellent agreement between the $\sim 100$ HDF 
spectroscopic redshifts and the predictions of the method, with a 
rms error $\Delta z/(1+z_{spec}) = 0.08$ up to $z<6$ and no 
systematic biases nor outliers. Note that these results have not been 
reached by minimizing the difference between spectroscopic and 
photometric redshifts (as is the case with empirical training set 
techniques), which may lead to an overestimation of the accuracy. The 
reliability of the method is further tested by restricting the color 
information to the UBVI filters. The results thus obtained are shown 
to be more reliable than those of standard techniques even when the 
latter include near-IR colors. 

The Bayesian formalism developed here can be generalized to deal 
with a wide range of problems which make use of photometric redshifts. 
Several applications are outlined, e.g. the estimation of individual 
galaxy characteristics as the metallicity, dust content, etc., or the 
study of galaxy evolution and the cosmological parameters from large 
multicolor surveys. Finally, using Bayesian probability it is possible to 
develop an integrated statistical method for cluster mass reconstruction 
which simultaneously considers the information provided by gravitational 
lensing and photometric redshift estimation.

\end{abstract} 


\section{Introduction}\label{int} 

The advent of the new class of 10-m ground based telescopes is 
having a strong impact on the study of galaxy evolution. For instance, 
instruments as LRIS at the Keck allow observers to regularly secure 
redshifts for dozens of $I\approx 24$ galaxies in several hours of exposure. 
Technical advances in the instrumentation, combined with the 
proliferation of similar telescopes 
in the next years guarantees a vast increase in the number of 
galaxies, bright and faint, for which spectroscopical redshifts will be 
obtained in the near future. Notwithstanding this progress 
in the sheer numbers of available spectra, the $I\approx 24$ `barrier' 
(for reasonably complete samples) is likely to stand for a time, 
as there are not foreseeable dramatic improvements in the telescope 
area or detection techniques. 

Despite the recent spectacular findings of very high redshift galaxies, 
(\cite{dey}, \cite{fra}, \cite{fry}), it is extremely difficult 
to secure redshifts for such objects. On the other hand, even 
moderately deep ground based imaging routinely contain many high 
redshift galaxies (although hidden amongst myriads of foreground ones), 
not to mention the Hubble Deep Field or the images that will be 
available with the upcoming Advanced Camera. To push further in redshift 
the study of galaxy evolution is therefore very important to develop 
techniques able to extract galaxy redshifts from multicolor photometry data. 

This paper applies the methods of Bayesian probability theory to 
photometric redshift estimation. Despite the efforts of 
Thomas Loredo, who has written stimulating reviews on the subject 
(Loredo 1990, 1992), Bayesian methods are still far from being one of 
the staple statistical techniques in Astrophysics. Most courses and 
monographs on Statistics only include a small section on Bayes' theorem, 
and perhaps as a consequence of that, Bayesian techniques are frequently 
used {\it ad hoc}, as another tool from the 
available panoply of statistical methods. However, as any reader of the 
fundamental treatise by E.T. Jaynes (1998) can learn, Bayesian probability 
theory represents an unified look to probability and statistics, which 
does not intend to complement, but to fully substitute the traditional, 
`frequentist' statistical techniques (see also Bretthorst 1988, 1990)

One of the fundamental differences between `orthodox' statistics and Bayesian 
theory, is that the probability is not defined as a frequency of occurrence, 
but as a reasonable degree of belief. Bayesian probability theory is 
developed as a rigorous full--flegded alternative to traditional 
probability and statistics based on this definition and 
three {\it desiderata}: a)Degrees of belief should be represented by 
real numbers, b)One should reason consistently, and c)The theory should 
reduce to Aristotelian logic when the truth values of hypothesis are known.

One of the most attractive features of Bayesian inference lies on its 
simplicity. There are two basic rules to manipulate probability, the 
product rule 
\begin{equation}
P(A,B|C)=P(A|C)P(B|A,C)
\end{equation}
and the sum rule
\begin{equation}
P(A+B|C)=P(A|C)+P(B|C)-P(A,B|C)
\end{equation}
where ``$A,B$'' means ``$A$ and $B$ are true'', and ``$A+B$'' 
means ``either $A$ or $B$ or both are true''. From the product rule, 
and taking into account that the propositions ``$A,B$'' and 
``$B,A$'' are identical, it is straightforward to derive Bayes' theorem 

\begin{equation}
P(A|B,C)={P(A|C)P(B|A,C)\over P(B|C)}
\end{equation}

If the set of proposals $B=\{ B_i\}$ are mutually exclusive and 
exhaustive, using the sum rule one can write
\begin{equation}
P(A,B|C)=P(A,\{B_i\}|C)=\sum_i P(A,B_i|C)
\label{mar}
\end{equation}
which is known as Bayesian marginalization. These are the basic 
tools of Bayesian inference. Properly used and combined with the 
rules to assign prior probabilities, they are in principle enough 
to solve most statistical problems. 

There are several differences between the methodology presented in 
this paper and that of \cite{kod}, the most significant being the 
treatment of priors (see Sec. \ref{bpz}). The procedures developed 
here offer a major improvement in the redshift estimation and based on them 
it is possible to generate new statistical methods for applications 
which make use of photometric redshifts (Sec. \ref{appli}). 

The outlay of the paper is the following: Sec. 2 reviews the current 
methods of photometric redshifts estimation making emphasis on 
their main sources of error. Sec. 3 introduces an expression 
for the redshift likelihood slightly different from the one used by other 
groups when applying the SED--fitting technique. In Sec. 4 it is 
described in detail how to apply Bayesian probability to photometric 
redshift estimation; the resulting method is called BPZ. Sec 5 
compares the performance of traditional statistical techniques, as 
maximum likelihood, with BPZ by applying both methods to the HDF 
spectroscopic sample and to a simulated catalog. Sec. 6 briefly 
describes how BPZ may be developed to deal with problems in galaxy 
evolution and cosmology which make use of photometric redshifts. 
Sec 7 briefly summarizes the main conclusions of the paper.

\section{Photometric redshifts: training set vs. SED fitting methods}
\label{sed}

There are two basic approaches to photometric redshift estimation. 
Using the terminology of \cite{yee}, they may be termed `SED fitting' 
and `empirical training set' methods. The first technique (\cite{koo}, 
\cite{lan}, \cite{gwy}, \cite{pel}, \cite{saw}, etc.) involves compiling 
a library of template spectra, empirical or generated with population 
synthesis techniques . These templates, after being redshifted and 
corrected for intergalactic extinction, are compared with the galaxy 
colors to determine the redshift $z$ which best fits the 
observations. The training set technique (\cite{bru}, \cite{con95}, \cite{wan}) 
starts with a multicolor galaxy sample with apparent 
magnitudes $m_0$ and colors $C$ which has been spectroscopically identified. 
Using this sample, a relationship of the kind $z=z(C,m)$ is determined using 
a multiparametric fit. 

It should be said that these two methods are more similar than what 
it is usually considered. To understand this, let's analyze 
how the empirical training set method works. For simplicity, 
let's forget about the magnitude dependence and let's suppose that only 
two colors $C=(C_1,C_2)$ are enough to estimate the photometric redshifts, 
that is, given a set of spectroscopic redshifts $\{z_{spec}\}$ and 
colors $\{C\}$, the training set method tries to fit a surface 
$z=z(C)$ to the data. It must be realized that this method makes a 
very strong assumption, namely that the surface $z=z(C)$ is a 
{\it function } defined on the color space: each value of $C$ is 
assigned one and only one redshift. Visually this means that 
the surface $z=z(C)$ does not `bend' over itself in the redshift 
direction. Although this functionality of the redshift/color 
relationship cannot be taken for granted in the general case 
(at faint magnitudes there are numerous examples of galaxies with 
very similar colors but totally different redshifts), it seems to be 
a good approximation to the real picture at $z<1$ redshifts and 
bright magnitudes (\cite{bru} ). A certain scatter around this surface 
is allowed: galaxies with the same value of $(C)$ may have slightly 
different redshifts and it seems to be assumed implicitly that this 
scatter is what limits the accuracy of the method. 

The SED fitting method is based on the color/redshift relationships 
generated by each of the library templates $T$, $C_T=C_T(z)$. 
A galaxy at the position $C$ is assigned the redshift corresponding 
to the closest point of any of the $C_T$ curves in the color space. 
If these $C_T$ functions are inverted, one ends up with the curves 
$z_T=z_T(C_T)$, which, in general, are not functions; they 
may present self--crossings (and of course they may also cross each other). 
If we limit ourselves to the region in the color/redshift space 
in which the training set method defines the surface $z=z(C)$,  
for a realistic template set the curves $z_T=z_T(C_T)$ would be 
embedded in the surface $z=z(C)$, conforming its `skeleton' and 
defining its main features. 

The fact that the surface $z=z(C)$ is continuous, whereas the 
template-defined curves are sparsely distributed, does not have a 
great practical difference. The gaps may be filled by finely interpolating 
between the templates (\cite{saw}), but this is not strictly necessary: 
usually the statistical procedure employed to search for the best redshift 
performs its own interpolation between templates. When the colors of a 
galaxy do not exactly coincide with one of the templates, $\chi^2$ or 
the maximum likelihood method will assign the redshift corresponding to 
the nearest template in the color space. 
This is equivalent to the curves $z_T=z_T(C_T)$ having extended 
`influence areas' around them, which conform a sort of step--like 
surface which interpolates across the gaps, and also extends beyond 
the region limited by them in the color space. Therefore, the SED-fitting 
method comes with a built-in interpolation (and extrapolation) procedure. 
For this reason, the accuracy of the photometric redshifts does not 
change dramatically when using a sparse template set as 
the one of \cite{cww} (\cite{lan}) or a fine grid of template spectra 
(\cite{saw}). The most crucial factor is that the template library, 
even if it contains few spectra, adequately reflects the main features 
of real galaxy spectra and therefore the main `geographical accidents' 
of the surface $z=z(C)$

The intrinsic similarity between both photometric redshift 
methods explains their comparable performance, especially at 
$z\lesssim 1$ redshift (\cite{hog}). When the topology of the 
color--redshift relationship is simple, as apparently happens at 
low redshift, the training set method will probably work slightly 
better than the template fitting procedure, if only because it avoids 
the possible systematics due to mismatches between the predicted 
template colors and the real ones, and also partially because it 
includes not only the colors of the galaxies, but also their 
magnitudes, what helps to break the color/redshift degeneracies 
(see below). However, it must be kept in mind that although 
the fits to the spectroscopic redshifts give only a dispersion 
$\delta z\approx 0.06$ (\cite{con97}), there is not a 
strong guarantee that the predictive capabilities of the training set 
method will keep such an accuracy, even within the same magnitude 
and redshift ranges. As a matter of fact, they do not seem to work 
spectacularly better than the SED fitting techniques (\cite{hog}), 
even at low and intermediate redshifts. 

However, the main drawback of the training set method is that, due to 
its empirical and {\it ad hoc} basis, in principle it can 
only be reliably extended as far as the spectroscopic redshift 
limit. Because of this, it may represent a cheaper method of 
obtaining redshifts than the spectrograph, but which cannot 
really go much fainter than it.
Besides it is difficult to transfer the information obtained with 
a given set of filters, to another survey which uses a different set. 
Such an extrapolation has to be done with the help of templates, what 
makes the method lose its empirical purity. And last but not least, 
it is obvious that as one goes to higher redshifts/fainter magnitudes 
the topology of the color-redshift distribution $z=z(C,m_0)$ displays 
several nasty degeneracies, even if the near-IR information is included, 
and it is impossible to fit a single functional form to the 
color-redshift relationship. 

Although the SED fitting method is not affected by some of these 
limitations, it also comes with its own set of problems. Several 
authors have analyzed in detail the main sources of errors affecting this 
method (\cite{saw},\cite{fsoto}). These errors may be divided into two 
broad classes:

\subsection {Color/redshift degeneracies} 

\begin{figure}[h]
\epsscale{0.5}
\plotone{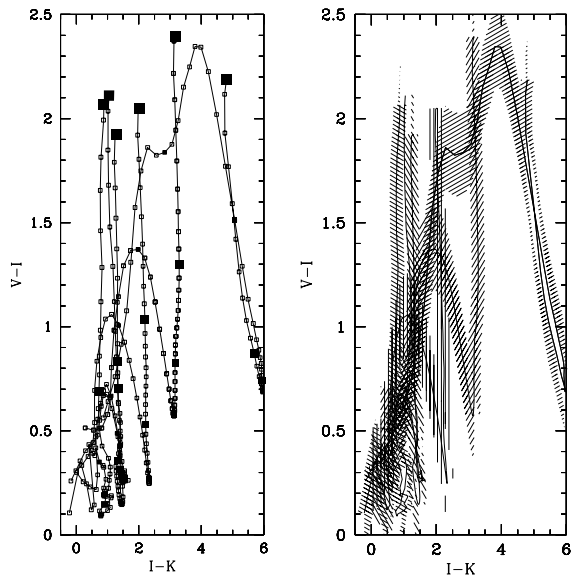}
\caption{a) On the left, VI vs. IK for the templates used 
in Sec \ref{test} in the interval $1<z<5$. The size of the filled squares 
grows with redshift, from $z=1$ to $z=5$. If these were the only colors 
used for the redshift estimation every crossing of the lines 
would correspond to a color/redshift degeneracy. b) To the right, 
the same color--color relationships `thickened' by a $0.2$ photometric 
error. The probability of color/redshift degeneracies highly increases.} 
\label{colors}
\end{figure}

Fig. \ref{colors}a. shows $VI$ vs $IK$ for the  morphological 
types employed in Sec \ref{test} and $0<z<5$. The color/redshift 
degeneracies happen when the line 
corresponding to a single template intersects itself or when two 
lines cross each other at points corresponding to different redshifts 
for each of them (these cases correspond to ``bendings'' in the 
redshift/color relationship $z=z(C)$). It is obvious that the likelihood 
of such crossings increases with the extension of the considered redshift 
range and the number of templates included. 

It may seem that even considering a very extended redshift range, such 
confusions could in principle be easily avoided by using enough 
filters. However, the presence of color/redshift 
degeneracies is highly increased by random photometric errors, 
which can be visualized as a blurring or thickening of the $C_T(z_T)$ 
relationship (fig. \ref{colors}b): each point of the curves in fig. 
\ref{colors}a is expanded into a square of size $\delta C$, the 
error in the measured color. The first consequence of this is a `continuous' 
($\delta z \approx { \partial C \over \partial z} \delta C$) increase in 
the rms of the `small-scale' errors in the redshift estimation, and, 
what it is worse, the overlaps in the color-color space become more 
frequent, with the corresponding rise in the number of `catastrophic' 
redshift errors. In addition, multicolor information may often 
be degenerate, so increasing the number of filters does not break 
the degeneracies; for instance, by applying a simple PCA analysis to the 
photometric data of the HDF spectroscopic sample it can be shown that 
the information contained in the seven $UBVIJHK$ filters for the HDF galaxies 
can be condensed using only three parameters, the coefficients of the 
principal components of the flux vectors (see also \cite{con95}). 
Therefore, if the photometric errors are large, it is not always possible 
to get totally rid of the degeneracies even increasing the number of 
filters. This means that the presence of color/redshift degeneracies is 
unavoidable for faint galaxy samples. The training set method somehow 
alleviates this problem by introducing an additional parameter in 
the estimation, the magnitude, which in some cases 
may break the degeneracy. However, it is obvious that color/redshift 
degeneracies also affect galaxies with the same magnitude, and the 
training set method does not even contemplate the possibility of their 
existence! 

The SED--fitting method at least allows for the existence 
of this problem, although it is not extremely efficient in dealing 
with it, especially with noisy data. Its choice of redshift is 
exclusively based on the goodness of fit between the observed 
colors and the templates. In cases as the one described above, where two 
or more redshift/morphological type combinations have practically the 
same colors, the value of the likelihood ${\mathcal L}$ would have two or more 
approximately equally high maxima at different redshifts (see 
fig. \ref{peaks}). Depending on the 
random photometric error, one maximum would prevail over the others, and a 
small change in the flux could involve a catastrophic change in 
the estimated redshift (see fig. \ref{peaks}). However, in many cases 
there is additional information, discarded by ML, which could potentially 
help to solve such conundrums. For instance, it may be known from previous 
experience that one of the possible redshift/type combinations is much 
more likely than any other given the galaxy magnitude, angular size, 
shape, etc. In that case, and since the likelihoods are not informative 
enough, it seems clear that the more reasonable decision would be to choose 
the option which is more likely {\it a priori} as the best estimate. Plain 
common sense dictates that one should compare all the possible hypotheses 
with the data, as ML does, but simultaneously keeping in mind the degrees of 
plausibility assigned to them by previous experience. There is not a 
simple way of doing this within ML, at best one may remove or change the 
redshift of the problematic objects by hand or devise {\it ad hoc} solutions 
for each case. In contrast, Bayesian probability theory allows to include 
this additional information in a rigorous and consistent way, effectively 
dealing with this kind of errors (see Sec \ref{bpz}) 

\subsection{Template incompleteness} 

In some cases, the spectra of observed galaxies have no close 
equivalents in the template library. Such galaxies will be 
assigned the redshift corresponding to the nearest template in 
the color/redshift space, no matter how distant from the observed color 
it is in absolute terms. The solution is obvious, one has to include 
enough templates in the library so that all the possible galaxy types 
are considered. 
As was explained above, the SED fitting techniques perform their own 
`automatic' interpolation and extrapolation, so once the main spectral 
types are included in the template library, the results are not greatly 
affected if one finely interpolates among the main spectra. The effects 
of using a correct but incomplete set of spectra are shown in Sec 
\ref{test}.

Both sources of errors described above are exacerbated at high redshifts. 
High redshift galaxies are usually faint, therefore with large photometric 
errors, and as the color/redshift space has a very extended range in $z$, 
the degeneracies are more likely; in addition the template incompleteness 
is worsened as there are few or no empirical spectra with which compare 
the template library. 

The accuracy of any photometric redshift technique is usually 
established by contrasting its output with a sample of galaxies with 
spectroscopic redshifts. It should be kept in mind, though, that 
the results of this comparison may be misleading, as the available 
spectroscopic samples are almost `by definition' especially well suited for 
photometric redshift estimation: relatively bright (and thus with small 
photometric errors) and often filling a privileged niche in the 
color-redshift space, far from degeneracies 
(e.g. Lyman-break galaxies). Thus, it is risky to extrapolate the 
accuracy reached by current methods as estimated from spectroscopic 
samples (and this also applies to BPZ) to fainter magnitudes. 
This is especially true for the training set methods, which deliberately 
minimize the difference between the spectroscopic and photometric 
redshifts. 

\section{Maximum likelihood (ML) redshift estimates}\label{ml}

Photometric redshift techniques based on template fitting look for the 
best estimate of a galaxy redshift from the comparison of its measured 
fluxes in $n_c+1$ filters $\{f_\alpha\}$, $\alpha=0,n_c$, with a set of 
$n_T$ template spectra which try to represent the different morphological 
types, and which have fluxes $f_{T\alpha}(z)$. These methods find their 
estimate $z_{ML}$ by maximizing the likelihood ${\mathcal L}$ (or 
equivalently minimizing $\chi^2$) over all the possible values of the 
redshift $z$, the templates $T$ and the normalization constant $a_0$. 
\begin{equation}
-\log({\mathcal L})+{\rm const} \propto \chi^2(z,T,a_0) = 
\sum_\alpha{(f_\alpha-a_0f_{T\alpha})^2\over 2\sigma_{f_\alpha}^2}
\label{li}
\end{equation}
Since the normalization constant $a_0$ is considered a free parameter, 
the only information relevant to the redshift determination is contained 
in the ratios among the fluxes $\{f_\alpha\}$, that is, in the galaxy colors. 

The definition of the likelihood in eq. (\ref{li}) is not convenient 
for applying Bayesian methods as it depends on a normalization 
parameter $a_0$, which is not convenient to define useful priors 
either theoretically or 
from previous observations. Here we prefer to normalize the 
total fluxes in each band by the flux in a `base' filter, e.g. 
the one corresponding to the band in which the galaxy sample 
was selected and is considered to be complete. 
Then the `colors' $C=\{c_i\}$, are defined as $c_i=f_i/f_0$ 
$i=1,n_c$, where $f_0$ is the base flux. The exact way in which the 
colors are defined is not relevant, other combinations of filters are 
equally valid. Hereinafter the magnitude $m_0$ (corresponding to the flux 
$f_0$) will be used instead of $f_0$ in the expressions for the 
priors. And so, assuming that the magnitude errors 
$\{\sigma_{f_\alpha}\}$ are gaussianly distributed, the likelihood can 
be defined as 

\begin{equation}
{\mathcal L}(T,z)=
{1 \over \sqrt{{(2\pi)}^{n_c}|\Lambda_{ij}|}}e^{-{\chi^2 \over 2}} 
\end{equation}
where
\begin{equation}
\chi^2=\sum_{i,j}\Lambda_{ij}^{-1}[c_i-c_{Ti}(z)][c_j-c_{Tj}(z)]
\end{equation}
and the matrix of moments $\Lambda_{ij}\equiv<\sigma_{c_i} \sigma_{c_j}>$ 
can be expressed as
\begin{equation}
\Lambda_{ij}=f_0^{-4}(f_i f_j \sigma_{f_0}^2 + f_0^2 
\delta_{ij}\sigma_{f_i}\sigma_{f_j})
\end{equation}

By normalizing by $f_0$ instead of $a_0$, one reduces the computational 
burden as it is not necessary to maximize over $f_0$, which is already 
the `maximum likelihood' estimate for the value of the galaxy flux in that 
filter. It is obvious that this assumes that the errors in the colors are 
gaussian, which in general is not the case, even if the flux errors are. 
Fortunately, the practical test performed below (Sec. \ref{test}) 
shows that there is little change between the results using both 
likelihood definitions (see fig. \ref{comparison}a).

\section{Bayesian photometric redshifts (BPZ)}\label{bpz}

Within the framework of Bayesian probability, the problem of photometric 
redshift estimation can be posed as finding the probability $p(z|D,I)$, 
i.e., the probability of a galaxy having redshift $z$ given the data 
$D=\{C,m_0\}$, {\it and} the prior information $I$. As it was mentioned 
in the introduction, Bayesian theory states that {\it all} the 
probabilities are conditional; they do not 
represent frequencies, but states of knowledge about hypothesis, and 
therefore always depend on other data or information (for a detailed 
discussion of this and many other interesting issues see Jaynes, 1998). 
The prior information $I$ is an ample term which in general 
should include any knowledge that may be relevant to the hypothesis 
under consideration and is not already included in the data $C,m_0$.
Note that in Bayesian probability the relationship between the prior 
and posterior information is {\it logical}; it does not have to be temporal 
or even causal. For instance, data from a new observation may 
be included as prior information to estimate the photometric redshifts of 
an old data set. Although some authors recommend that the  $`|I'$ should not 
be dropped from the expressions of probability (as a remainder of the fact 
that all probabilities are conditional and especially to avoid confusions 
when two probabilities based on different prior informations are considered 
as equal), here the rule of simplifying the mathematical notation 
whenever there is no danger of confusion will be followed, and from 
now $p(z)$ will stand for $p(z|I)$, $p(D|z)$ for $p(D|z,I)$ etc. 

As a trivial example of the application of Bayes's theorem, let's 
consider the case if which there is only one template and the 
likelihood ${\mathcal L}$ only depends on the redshift 
$z$. Then, applying Bayes theorem 
\begin{equation} 
p(z|C,m_0)={p(z|m_0) p(C|z) \over p(C)} \propto p(z|m_0)p(C|z) 
\label{1}
\end{equation} 
The expression $p(C|z)\equiv {\mathcal L}$ is simply the likelihood: 
the probability of observing the colors $C$ if the galaxy 
has redshift $z$ (it is assumed for simplicity that ${\mathcal L}$ only 
depends on the redshift and morphological type, and not on $m_0$)  
The probability $p(C)$ is a normalization constant, and usually 
there is no need to calculate it. 

The first factor, the {\it prior} probability $p(z|m_0)$, is the 
redshift distribution for galaxies with magnitude $m_0$. This function 
allows to include information as the existence of upper or lower limits 
on the galaxy redshifts, the presence of a cluster in the field, etc. 
The effect of the prior $p(z|m_0)$ on the estimation depends on how 
informative it is. It is obvious that for a constant prior 
(all redshifts equally likely {\it a priori}) the estimate obtained from 
eq. (\ref{1}) will exactly coincide with the ML result. This is also 
roughly true if the prior is `smooth' enough and does not 
present significant structure. However, in other cases, values of 
the redshifts which are considered very improbable from the prior 
information would be ``discriminated''; they must fit the data much 
better than any other redshift in order to be selected. 

Note that in rigor, one should write the prior in eq. (\ref{1}) as  
\begin{equation} 
p(z|m_0)
=\int d\hat{m_0}p(\hat{m_0})p(m_0|\hat{m_0})p(z|\hat{m_0})
\end{equation} 
where $\hat{m_0}$ is the `true' value of the observed magnitude $m_0$, 
$p(\hat{m_0})$ is proportional to the number counts as a function 
of the magnitude $m_0$ and $p(m_0|\hat{m_0})\propto 
\exp[(m_0-\hat{m_0})^2/2\sigma_{m_0}^2]$, i.e, the probability of 
observing $m_0$ if the true magnitude is $\hat{m_0}$. 
The above convolution accounts for the uncertainty in 
the value of the magnitude $m_0$, which has the effect of slightly 
`blurring' and biasing the redshift distribution $p(z|m_0)$. 
To simplify our exposition this effect would not be consider hereinafter, 
and just $p(z|m_0)$ and its equivalents will be used.

\subsection{Bayesian Marginalization}

It may seem from eq. \ref{1} (and unfortunately it is quite a 
widespread misconception) that the only difference between Bayesian 
and ML estimates is the introduction of a prior, in this case, 
$p(z|m_0)$. However, there is more to Bayesian probability than just 
priors. 

The galaxy under consideration may belong 
to different morphological types represented by a set of $n_T$ templates. 
This set is considered to be {\it exhaustive}, i.e including all possible 
types, and {\it exclusive}: the galaxy cannot belong to two types at 
the same time. In that case, using Bayesian marginalization (eq. \ref{mar}) 
the probability $p(z|D)$ can be `expanded' into a `basis' formed by the 
hypothesis $p(z,T|D)$ (the probability of the galaxy redshift 
being $z$ {\it and} the galaxy type being $T$). The sum over all 
these `atomic' hypothesis will give the total probability $p(z|D)$. 
That is, 
\begin{equation}
p(z|C,m_0)=\sum_T p(z,T|C,m_0)\propto 
\sum_T p(z,T|m_0)p(C|z,T)
\label{bas}
\end{equation}
$p(C|z,T)$ is the likelihood of the data $C$ given $z$ and $T$. 
The prior $p(z,T|m_0)$ may be developed using the product rule. 
For instance
\begin{equation}
p(z,T|m_0)=p(T|m_0)p(z|T,m_0)
\label{pri}
\end{equation}
where $p(T|m_0)$ is the galaxy type fraction as a function 
of magnitude and $p(z|T,m_0)$ is the redshift distribution for 
galaxies of a given spectral type and magnitude.

\begin{figure}[h]
\epsscale{0.5}
\plotone{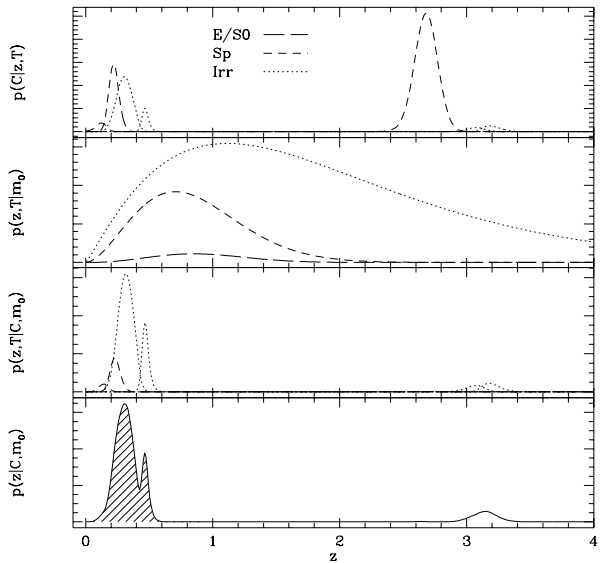}
\caption{An example of the main probability 
distributions involved in BPZ for a galaxy at $z=0.28$ with 
an Irr spectral type and $I\approx 26$ to which random photometric 
noise is added. From top 
to bottom: a) The likelihood functions $p(C|z,T)$ for the different 
templates used in Sec \ref{test}. Based on ML, the redshift chosen for 
this galaxy would be $z_{ML}=2.685$ and its spectral type would correspond 
to a Spiral. b) The prior probabilities $p(z,T|m_0)$ for each of the 
spectral types (see text). Note that the probability of finding 
a Spiral spectral type with $z>2.5$ and a magnitude $I=26$ is almost 
negligible. c) The probability distributions 
$p(z,T|C,m_0)\propto p(z,T|m_0)p(C|z,T)$ , that is, the likelihoods in 
the top plot multiplied by the priors. The high redshift peak due to 
the Spiral has disappeared, although there is still a little chance of 
the galaxy being at high redshift if it has a Irr spectrum, but the 
main concentration of probability is now at low redshift.
d) The final Bayesian probability $p(z|C,m_0)=\sum_T p(z,T|C,m_0)$, 
which has its maximum at $z_B=0.305$. The shaded area corresponds to 
the value of $p_{\Delta z}$, which estimates the reliability of $z_B$ and 
yields a value of $\approx 0.91$.}
\label{peaks}
\end{figure}

Eq. (\ref{bas}) and fig. \ref{peaks} clearly illustrate the main 
differences between the 
Bayesian and ML methods. ML would just pick the 
highest maximum over all the $p(C|z,T)$ as the best redshift estimate, 
without looking at the plausibility of the corresponding values of 
$z$ or $T$. On the contrary, Bayesian probability averages all these 
likelihood functions after weighting them by their prior probabilities 
$p(z,T|m_0)$. In this way the estimation is not affected by spurious 
likelihood peaks caused by noise as it is shown in fig. 
\ref{peaks} (see also the results of Sec. \ref{test}). Of course 
that in an ideal situation with perfect, noiseless observations (and a 
nondegenerate template space, i.e, only one $C$ for each $(z,T)$ pair) 
the results obtained with ML and Bayesian inference would be the same.

Instead of a discrete set of templates, the comparison library may contain 
spectra which are a function of continuous parameters. For instance, 
synthetic spectral templates depend on the metallicity $Z$, the dust 
content, the star formation history, etc. Even starting from a set of a 
few templates, they may be expanded using the principal component analysis 
(PCA) technique (\cite{sod}). In general, 
if the spectra are characterized by $n_A$ possible parameters 
$A=\{a_1...a_{n_A}\}$ (which may be physical characteristics of the models 
or just PCA coefficients), the probability of $z$ given $F$ can be expressed 
as 
\begin{equation}
p(z|C,m_0)=\int dA p(z,A|C,m_0)
\propto\int dA p(z,A|m_0)p(C|z,A)
\label{cont}
\end{equation}

\subsection{`Bookmaker' odds}

Sometimes, instead of finding a `point' estimate for a galaxy redshift, 
one needs to establish if that redshift belongs within a certain interval. 
For instance, the problem may be to determine whether the galaxy has 
$z>z_t$, where $z_t$ is a given threshold, or whether its redshift falls 
within a given $z \pm \Delta z$, e.g. in the selection 
of cluster members or background galaxies for lensing studies.  

As an example, let's consider the classification of galaxies into the 
background-foreground classes with respect to a certain redshift 
threshold $z_{th}$. 
One must choose between the hypothesis $H_{th}=\{ z > z_{th}\}$ and 
its opposite, $\bar{H}_{th}=\{ z < z_{th}\}$. The corresponding 
probabilities may be written as 
\begin{equation}
P(H_{th}|D)=\int_0^{z_{th}}dz p(z|D)  
\end{equation}
And 
\begin{equation}
P(\bar{H}_{th}|D)=\int_{z_{th}}^{\infty}dz p(z|D)  
\end{equation}
The (`bookmaker') odds of hypothesis $H_{th}$ are defined as the probability 
of $H_{th}$ being true over the probability of $H_{th}$ being false (Jaynes 1998)
\begin{equation}
O(H_{th}|D)={P(H_{th}|D)\over P(\bar{H}_{th}|D)}
\end{equation}
When $O(H_{th}|D)\approx 1$, there is not enough information to choose 
between both hypothesis. A galaxy is considered to have $z>z_{th}$ if 
$O(H_{th}|D)>O_{d}$, where $O_{d}$ is a certain decision threshold. There 
are no fixed rules to choose the value of $O_{d}$, and the most 
appropriate value depends on 
the task at hand; for instance, to be really sure that no foreground 
galaxy has sneaked into the background sample, $O_{d}$ would have to be 
high, but if the main goal is selecting all the background galaxies and 
one does not mind including some foreground ones, then $O_{d}$ would be 
lower, etc. Basically this is a problem concerning decision theory.

In the same way, the cluster galaxies can be selected by choosing a 
redshift threshold $\Delta z$ which defines whether a galaxy belongs 
to the cluster. The corresponding hypothesis would be 
$H_c=\{|z-z_c| < \Delta z\}$. 
\begin{equation}
P(H_c|D)=\int_{z_c-\Delta z}^{z_c+\Delta_z} dz p(z|D)  
\end{equation}
And 
\begin{equation}
P(\bar{H}_c|D=\int_0^{z_c-\Delta z}dz p(z|D)+ 
\int_{z_c+\Delta z}^{\infty} dz p(z|D)
\end{equation}
Similarly, the odds of $H_c$ are defined as 
\begin{equation}
O(H_c|D)={P(H_c|D)\over P(\bar{H}_c|D}
\end{equation}

\subsection{Prior calibration}

In those cases where the prior information is vague and 
does not allow to choose a definite expression prior probability, 
Bayesian inference offers the possibility of ``calibrating'' the prior, 
if needed using the very data sample under consideration. 

Let's suppose that the distribution $p(z,T,m_0)$ is parametrized using 
$n_\lambda$ continuous parameters $\lambda$. They may be the coefficients 
of a polynomial fit, a wavelet expansion, etc. In that case, including 
$\lambda$ in eq. (\ref{bas}), the probability can be written as 
\begin{equation}
p(z|C,m_0)=\int d\lambda\sum_T p(z,T,\lambda|C,m_0)\propto
\int d\lambda p(\lambda)\sum_T p(z,T,m_0|\lambda)p(C|z,T)
\label{20}
\end{equation}
where $p(\lambda)$ is the prior probability of $\lambda$, and 
$p(z,T,m_0|\lambda)$ is the prior probability of $z,T$ and $m_0$ 
as a function of the parameters $\lambda$. The latter have not been 
included in the likelihood expression since $C$ is totally determined 
once the values of $z$ and $T$ are known. 

Now let's suppose that the galaxy belongs to a sample 
containing $n_g$ galaxies. Each $j-$th galaxy has a `base' magnitude 
$m_{0j}$ and colors $C_j$. The sets ${\bf C}\equiv\{C_j\}$ and 
${\bf m_0}\equiv \{m_{0j}\}, j=1,n_g$ contain respectively the colors 
and magnitudes of all the galaxies in the sample. 
Then, the probability of the $i-$th galaxy 
having redshift $z_i$ given the full sample data 
${\bf C}$ and ${\bf m_0}$ can be written as
\begin{equation}
p(z_i|{\bf C},{\bf m_0})=
\int d\lambda\sum_{T} p(z_i,T,\lambda|C_i,m_{0i},{\bf C'},{\bf m_0'})
\end{equation}
The sets ${\bf C'}\equiv \{C_j\}$ and ${\bf m_0'}\equiv\{m_{0j}\}$, 
$j=1,n_g, j\neq i$ are identical to ${\bf C}$ and ${\bf m_0}$ 
except for the exclusion of the data $C_i$ and $m_{0i}$. 
Applying Bayes' theorem, the product rule and simplifying 
\begin{equation}
p(z_i|{\bf C},{\bf m_0})\propto
\int d\lambda p(\lambda|{\bf C'},{\bf m_0'})
\sum_{T} p(z_i,T,m_{0i}|\lambda)p(C|z_i,T)
\end{equation}
where as before it has been considered that the likelihood of $C_i$ 
only depends on $z_i,T$  and that the probability of $z_i$ and $T$ 
only depend on ${\bf C'}$ and ${\bf m_0'}$  through $\lambda$. 
The expression to which we arrived is very similar to eq. (\ref{20}) 
only that now the shape of the prior is estimated from 
the data ${\bf C'},{\bf m_0'}$. This means that even if one starts with 
a very sketchy idea about the shape of the prior, the very galaxy sample 
under study can be used to determine the value of the parameters $\lambda$, 
and thus to provide a more accurate estimate of the individual galaxy 
characteristics. Assuming that the data ${\bf C'}$ (as well as 
${\bf m_0}$) are independent among themselves 
\begin{equation}
p(\lambda|{\bf C'},{\bf m_0'})\propto p(\lambda)
p({\bf C'},{\bf m_0'}|\lambda)= p(\lambda)
\prod_{j,j\neq i} p(C_j,m_{0j}|\lambda) 
\label{23}
\end{equation}
where
\begin{equation}
p(C_j,m_{0j}|\lambda)= \int dz_j\sum_{T_j} p(z_j,{T_j},C_j,m_{0j}|\lambda) 
\propto
\int dz_j\sum_{T_j} p(z_j,{T_j},m_{0j}|\lambda)p(C|z_j,T_j)
\end{equation} 

If the number of galaxies in our sample is large enough, it can be 
reasonably assumed that the prior probability 
$p(\lambda|{\bf C'},{\bf m_0'})$ will not 
change appreciably with the inclusion of the data $C_i,m_{0i}$ 
belonging to a single galaxy. In that case, a time-saving approximation 
is to use as a prior the probability $p(\lambda|{\bf C},{\bf m_0})$, 
calculated using the whole data set, instead of finding 
$p(\lambda|{\bf C'},{\bf m_0'})$ for each galaxy. In addition, 
it should be noted that $p(\lambda|{\bf C},{\bf m_0})$ represents 
the Bayesian estimate of the parameters which define the shape of 
the redshift distribution (see fig. \ref{nz}).  

\subsection{Including spectroscopical information}

In some cases spectroscopical redshifts $\{z_{si}\}$ 
are available for a fraction of the galaxy sample. It is 
straightforward to include them 
in the prior 
calibration procedure described above, using a delta--function 
likelihood weighted by the probability of the galaxy belonging to a 
morphological type, as it is done to determine the priors 
in Sec \ref{test}. This gives the spectroscopical subsample a 
(deserved) larger weight in the determination of the redshift and 
morphological priors in comparison with the rest of the galaxies, at 
least within a certain color and magnitude region, but, unlike what 
happens with the training set method, the information contained in the 
rest of the sample is not thrown away.
 
If nevertheless one wants to follow the training set approach and 
use only the spectroscopic sample, it is easy to develop a Bayesian 
variant of this method. As before, the 
goal is to find an expression of the sort $p(z|C,m_0)$, which 
would give us the redshift probability for a galaxy given its colors and 
magnitude. If the color/magnitude/redshift multidimensional 
surface $z=z(C,m_0)$ were infinitely thin, the probability 
would just be $p(z|C,m_0)\equiv \delta(z(C,m_o))$, where $\delta(...)$ 
is a delta-function. But in the real world there is always some 
scatter around the surface defined by $z(C,m_0)$ (even without taking 
into account the color/redshift degeneracies), and it is therefore 
more appropriate to describe $p(z|C,m_0)$ as e.g. a gaussian of 
width $\sigma z$ centered on each point of the surface $z(C,m_0)$. 
Let's assume that all the parameters which define the shape of this
relationship, together with $\sigma z$ are included in the set $\lambda_z$. 
Using the prior calibration method introduced above, the probability 
distribution for these parameters $p(\lambda_z|D_T)$ can be 
determined from the training set $D_T\equiv \{z_{si},Ci,m_{0i}\}$.
\begin{equation}
p(\lambda_z|D_T)\propto p(\lambda_z) 
\prod_i p(z_{si}|C_i,m_{0i},\lambda_z)
\end{equation}

The expression for the redshift probability of a galaxy with colors 
$C$ and $m_0$ would then be
\begin{equation}
\label{25}
p(z|C,m_0)=\int d\lambda_z p(\lambda_z|D_T)p(z|C,m_0,\lambda_z) 
\end{equation}

 The redshift probability obtained from eq. (\ref{25}) is compatible with 
the one obtained in eq. (\ref{bas}) using the SED--fitting 
procedure. Therefore it is possible to combine them 
in a same expression. As an approximation, let's suppose that 
both of them are given equal weights, then 
\begin{equation}
p(z|C,m_0)\propto \sum_T p(z,T|m_0)p(C|z,T)
+\int d\lambda_z p(\lambda_z|D)p(z|C,m,\sigma_z,\lambda_z) 
\end{equation}

In fact, due to the above described redundancy 
between the SED--fitting method and the training set method (Sec. 
\ref{sed}), it 
would be more appropriate to combine both probabilities using 
weights which would take these redundancies into account in a 
consistent way, roughly using eq.(\ref{25}) at brighter magnitudes, 
where the galaxies are well studied spectroscopically and 
leaving eq.(\ref{bas}) for fainter magnitudes. The exploration of 
this combined, training set/SED-fitting approach will be left for 
a future paper, and in the practical tests performed below the 
followed procedure uses the SED--fitting likelihood. 

\section{A practical test for BPZ}\label{test}
\begin{figure}[h]
\epsscale{0.5}
\plotone{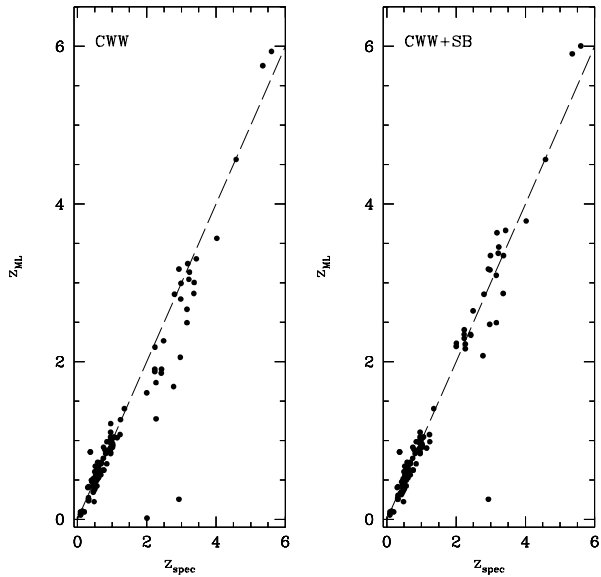}
\caption{a)To the left, the photometric redshifts obtained by applying our 
ML algorithm to the HDF spectroscopic sample using a template library which 
contains only the four CWW main types, E/SO, Sbc, Scd and Irr. 
These results are very similar to those of Fern\'andez-Soto, 
Lanzetta \& Yahil, 1998. b)
The right plot shows the significant improvement (without using BPZ yet) 
obtained by just including two of the Kinney et al. 1996 spectra of 
starburst galaxies, SB2 and SB3, in the template set. One of the outliers 
disappears, the `sagging' or systematic offset between $1.5<z<3.5$ is 
eliminated and the general scatter of the relationship decreases from 
$\Delta z/(1+z_{spec})=0.13$ to $\Delta z/(1+z_{spec})=0.10$.}
\label{comparison}
\end{figure}

The Hubble Deep Field (HDF; \cite{wil}) has become {\it the} 
benchmark in the development of photometric redshift techniques. 
In this section BPZ will be applied to the HDF and 
its performance contrasted with the results obtained with the standard 
`frequentist' (in the Bayesian terminology) method, the procedure 
usually applied to the HDF (\cite{gwy},\cite{lan},
\cite{saw}, etc.). The photometry used for the HDF is that of 
\cite{fsoto}, which, in addition to magnitudes in the four 
HDF filters includes JHK magnitudes from the observations of \cite{dic}. 
$I_{814}$ is chosen as the base magnitude. The colors are defined as 
described in Sec. \ref{ml}.

The template library was selected after several tests with the HDF 
subsample which has spectroscopic redshifts (108 galaxies), spanning the 
range $z<6$. The set of spectra which worked best is similar to that 
used by \cite{saw}. It contains four \cite{cww} 
templates (E/S0, Sbc, Scd, Irr), that is the same spectral types used 
by \cite{fsoto}, plus the spectra of 2 starbursting galaxies from 
\cite{kin}(\cite{saw} used two very blue SEDs from GISSEL). 
All the spectra were extended to the UV using a linear extrapolation 
and a cutoff at $912\AA$, and to the near--IR using GISSEL synthetic 
templates. The spectra are corrected for intergalactic absorption 
following \cite{mad}.

\begin{figure}[h]
\epsscale{0.5}
\plotone{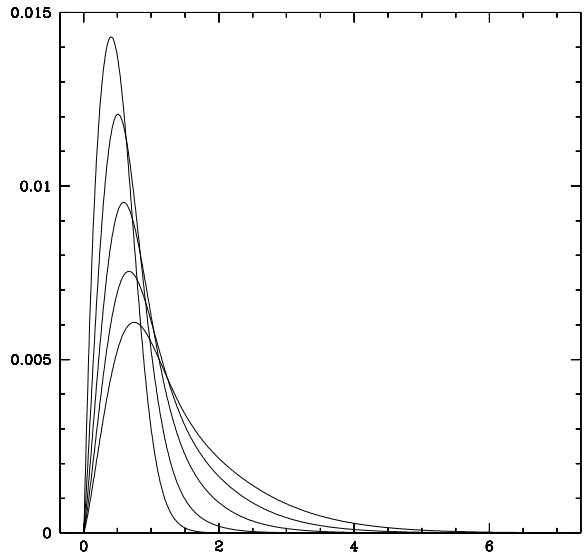}
\caption{The prior in redshift $p(z|m_0)$ estimated from the HDF 
data using the prior calibration procedure described in Sec 4., 
for different values of the magnitude $m_0$ ($I_{814}=21$ to 
$I_{814}=28$)}
\label{priors}
\end{figure}

It could seem in principle that a synthetic template set which takes 
(at least tentatively) into account galaxy evolution is more appropriate 
than `frozen' template library obtained at low redshift and then 
extrapolated to very high redshifts. However, as 
\cite{yee} has convincingly shown, the extended CWW set offers much 
better results than the GISSEL synthetic models \cite{bc}. 
I have also tried to use the RVF set of spectra and the agreement 
with the spectroscopic redshifts is considerably worse than using 
the empirical template set. And if the synthetic models do not work well 
within the magnitude range corresponding to the HDF spectroscopic sample 
is relatively bright, there is little reason to suppose that their 
performance will improve at fainter magnitudes.

However, even working with empirical templates, it is important 
to be sure that the template library is complete enough. 
Fig. \ref{comparison} illustrates the effects of template incompleteness 
in the redshift estimation. The left plot displays the results obtained 
using ML (Sec \ref{ml}) redshift estimation using only the four 
CWW templates (this plot is very similar to the $z_{phot}-z_{spec}$ 
diagram shown in \cite{fsoto}, which confirms the validity of the 
expression for the likelihood introduced in Sec \ref{ml}). 
On the right, the results obtained also using ML (no BPZ yet) but 
including two more templates, SB2 and SB3 from \cite{kin}. It can be 
seen that the new templates almost do not affect the low redshift range, 
but the changes at $z>2$ are quite dramatic, the `sagging' of the 
CWW--only diagram disappears and the general scatter of the diagram 
decreases by $20\%$. This shows how important it is to include enough 
galaxy types in the template library. No matter how sophisticated 
the statistical treatment is, it will do little to improve the results 
obtained with a deficient template set.
  
  The first step in the application of BPZ is choosing the shape of 
the priors. Due to the depth of the HDF there is little previous 
information about the redshift priors, so this is a good example in 
which the prior calibration procedure described in Sec\ref{bpz} has 
to be applied. It will be assumed that the early types (E/S0) 
and spirals (Sbc,Scd) have a spectral type prior (eq. \ref{pri} ) 
of the form
\begin{equation}
p(T|m_0)=f_t e^{-k_t (m_0-20)} 
\label{par}
\end{equation}
with $t=1$ for early types and $t=2$ for spirals. 
The irregulars (the remaining three templates; $t=3$) complete 
the galaxy mix. The fraction of early types at 
$I=20$ is assumed to be $f_1=35\%$ and that of spirals $f_2=50\%$. 
The parameters $k_1$ and $k_2$ are left as free.  
Based on the result from redshift surveys
the following shape for the redshift prior has been chosen:
\begin{equation}
p(z|T,m_0)\propto z^{\alpha_t} 
exp
\{ 
-[{z \over z_{mt}(m_o)}]^{\alpha_t} 
\}
\label{par1}
\end{equation}
where 
\begin{equation}
z_{mt}(m_0)=z_{0t}+k_{mt} (m_0-20.)
\label{par2}
\end{equation}
and $\alpha_t$, $z_{0t}$ and $k_t$ are considered free parameters. 
In total, 11 parameters have to be determined using the 
calibration procedure. For those objects with spectroscopic 
redshifts, a `delta-function' located at the spectroscopic redshift 
of the galaxy has been used instead of the likelihood $p(C|z,T)$.
Table 1 shows the values of the `best' values of the parameters 
in eq. (\ref{par1},\ref{par2}) found by maximizing the 
probability in eq. (\ref{25}) using the subroutine {\it amoeba} 
(\cite{nr}). 
The errors roughly indicate the parameter range which encloses $66\%$ 
of the probability. The values of the parameters in eq. (\ref{par}) 
are $k1=0.47\pm 0.02$ and $k2=0.165\pm 0.01$. The prior in 
redshift $p(z|m_0)$ can obviously be found by summing over 
the `nuisance' parameter (Jaynes 1998), in this case $T$:
\begin{equation}
p(z|m_0)=\sum_T p(T|m_0)p(z|T,m_0)
\end{equation}
Fig. \ref{priors} plots this prior for different magnitudes. 

With the priors thus found, one can proceed with the redshift 
estimation using eq. (\ref{20}). Here the multiplication by 
the probability distribution $p(\lambda)$ and the integration 
over $d \lambda$ will be skipped. As it can be seen from Table 1, 
the uncertainties in the parameters are rather small and it is obvious 
that the results would not change appreciably, so the additional 
computational effort of performing a 11-dimensional integral is not 
justified. 

There are several options to convert the continuous probability 
$p(z|C,m_0)$ to a point estimate of the `best' redshift $z_{B}$.  
Here the `mode' of the final probability is chosen, although 
taking the `median' value of $z$, corresponding to $50\%$ of the 
cumulative probability, or even the `average' 
$<z>\equiv\int dz z p(z|C,m_0)$ is also valid. 

\begin{figure}[h]
\epsscale{0.5}
\plotone{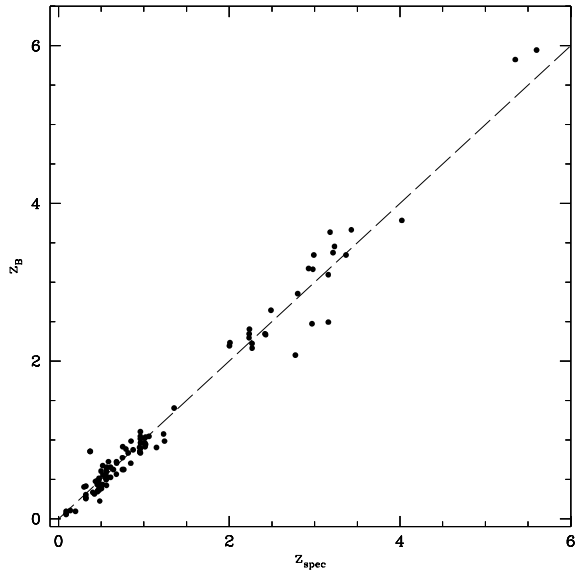}
\caption{The photometric redshifts obtained with BPZ plotted against 
the spectroscopic redshifts. The differences with fig. 
\ref{comparison}b are the elimination of 3 galaxies with 
$p_{\Delta z}<0.99$ (see text). This removes the only outlier present 
in fig. \ref{comparison}b. The rms scatter around the continuous line 
is $\Delta z_B/(1+z_B)=0.08$. }
\label{hdf}
\end{figure}

It was mentioned in sec \ref{bpz} that bayesian probability 
offers a way to characterize the accuracy of the redshift 
estimation using the odds or a similar indicator, for 
instance by analogy with the gaussian distribution a 
`$1\sigma$' error may be defined using a interval with contains 
$66\%$ of the integral of $p(z|C,m_0)$ around $z_{B}$, etc. 
Here it has been chosen as an indicator of the redshift reliability 
the quantity $p_{\Delta z}$, the probability of $|z-z_B|<\Delta z$, 
where $z$ is the galaxy redshift. In this way, when the value of 
$p_{\Delta z}$ is low, we are warned that the redshift prediction is 
unreliable. As it will be shown below, $p_{\Delta z}$ is extremely 
efficient in picking out galaxies with `catastrophic errors' in their 
redshifts. 
  The photometric redshifts resulting from applying BPZ to the 
spectroscopic sample are plotted in fig. \ref{hdf}. Galaxies with 
a probability $p_{\Delta z}<0.99$ (there are three of them) have 
been discarded, where $\Delta z$ is chosen to be $0.2\times(1+z)$, 
to take into account that the uncertainty grows with the redshift of 
the galaxies. 

It is evident from fig. \ref{hdf} that the agreement is very good 
at all redshifts. The residuals $\Delta z_B=z_{B}-z_{spec}$ 
have $<\Delta z_B>=0.002$. If $\Delta z_B$ is divided by a 
factor $(1+z_{spec})$, as suggested in \cite{fsoto}, the rms 
of the quantity $\Delta z_B/(1+z_B)$ is only $0.08$. There are no 
appreciable systematic effects in the residuals. One of 
the three objects discarded because of their having $p_{\Delta z}<0.99$ 
is the only clear outlier in our ML estimation, with $z_{BPZ}=0.245$ and 
$z_{spec}=2.93$ (see fig. \ref{comparison}b), evidence of the 
usefulness of $p_{\Delta z}$  to generate a reliable sample. 

From the comparison of fig. \ref{comparison}b with fig. 
\ref{hdf}, it may seem that, apart from the exclusion of the 
outlier, there is not much profit in applying BPZ with respect to ML. 
This is not surprising in the particular case of the HDF 
spectroscopic sample, which is formed mostly by galaxies either 
very bright or occupying privileged regions in the color space. The 
corresponding likelihood peaks are thus rather sharp, and little 
affected by smooth prior probabilities. 

\begin{figure}[h]
\epsscale{0.5}
\plotone{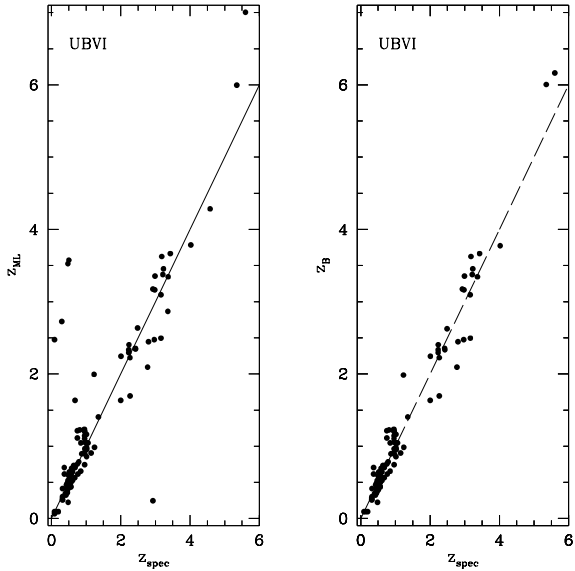}
\caption{a) The left plot shows the results of applying ML to the HDF 
spectroscopic sample using only the four HST bands. Compare with fig. 
\ref{comparison}b, which uses also the near IR photometry of 
Dickinson et al. 1998. 
The rms of the diagram is increased, and there are several outliers. 
b) The right plot shows how applying BPZ with a threshold 
$p_{\Delta z}>0.99$ leaves the remaining 101 objects ($93.5\%$ of the total)
virtually free of outliers. It is noteworthy that these results are 
totally comparable or even better (as there are no outliers) than  
those obtained in fig. \ref{comparison}a, in which the near-IR magnitudes 
were included in the estimation. 
}
\label{4f}
\end{figure}

To illustrate the effectiveness of BPZ under worse than ideal 
conditions, the photometric redshifts for the spectroscopic 
sample are estimated again using ML and BPZ but restricting the 
color information to the UBVI HST filters.
The results are plotted in fig. \ref{4f}. The ML redshift diagram 
displays 5 `catastrophic errors' ($\Delta z\gtrsim 1$). Note 
that these are the same kind of errors pointed out by \cite{ellis} 
in the first HDF photometric redshifts estimations. 
BPZ with a $p_{\Delta z}>0.99$ threshold (which eliminates a total of 7 
galaxies) totally eliminates those outliers. This is a clear example 
of the capabilities of BPZ (combined with an adequate template set) 
to obtain reliable photometric redshift estimates. Note that even using 
near--IR colors, the ML estimates shown in fig. \ref{comparison} presented 
outliers. This shows that applying BPZ to UV--only data may yield results 
more reliable than those obtained with ML including near-IR 
information! Although of course no more accurate; the scatter of 
fig. \ref{comparison}b, once the outliers are removed is 
$\Delta z\approx 0.18$, whereas fig. \ref{4f}b has a scatter of 
$\Delta z\approx 0.24$, which incidentally is approximately the scatter 
of fig. \ref{comparison}a. 

\begin{figure}[h]
\epsscale{0.5}
\plotone{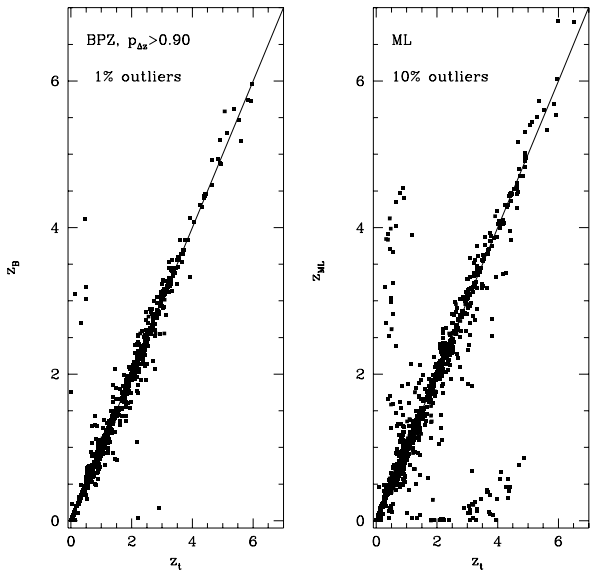}
\caption{a) To the left, the photometric redshifts $z_B$ 
estimated using BPZ for the $I<28$ HDF mock catalog, plotted against 
the `true' redshifts $z_t$ (see text). A threshold of $p_{\Delta z}>0.90$, 
which eliminates $20\%$ of the objects has been applied. 
b) The right plot shows the results obtained applying ML to the same mock 
sample. The fraction of outliers is $10\%$).}
\label{90}
\end{figure}

Another obvious way of testing the efficiency of BPZ is with a 
simulated sample. The latter can be generated using the procedure 
described in \cite{fsoto}. Each galaxy in the 
HDF is assigned a redshift and type using 
ML (this is done deliberately to avoid biasing the test 
towards BPZ) and then a mock catalog is created containing the colors 
corresponding to the best fitting redshifts and templates. To represent
the photometric errors present in observations, a random photometric 
noise of the same amplitude as the photometric error is added to each 
object. Fig. \ref{90}b shows the ML estimated redshifts for the mock 
catalog ($I<28$) against the `true' redshifts; although in general 
the agreement is not bad (as could be expected) there are a large number 
of outliers ($10\%$), whose positions illustrate the main source 
of color/redshift degeneracies: high $z$ galaxies 
which are erroneously assigned $z\lesssim 1$ redshifts and vice versa. 
This shortcoming of the ML method is analyzed in detail in \cite{fsoto}.
In contrast, fig. \ref{90}a shows the results of applying BPZ with 
a threshold of $p_{\Delta z}>0.9$. This eliminates $20\%$ of the 
initial sample (almost half of which have catastrophically wrong 
redshifts), but the number of outliers is reduced to a remarkable 
$1\%$.

\begin{figure}[h]
\epsscale{0.5}
\plotone{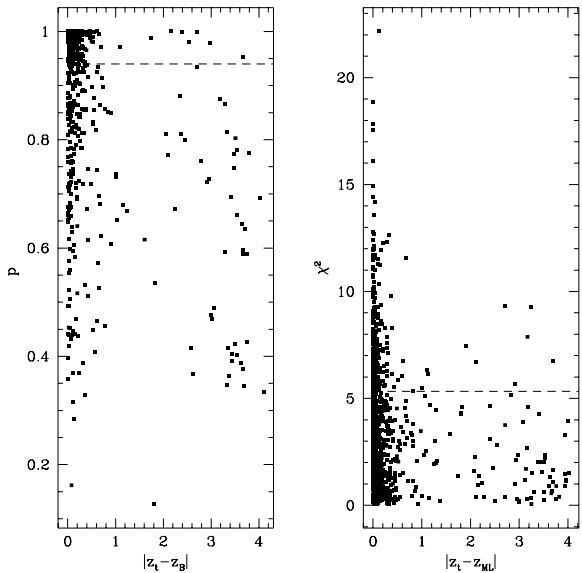}
\caption{a) On the left, the probability $p_{\Delta z}$ plotted 
against the absolute value of the difference between the `true' 
redshift ($z_t$) and the one estimated using BPZ ($z_B$) for the mock 
sample described in Sec. \ref{test}. The higher the value 
of $p_{\Delta z}$, the more reliable the redshift should be. The dashed 
line shows the $25\%$ low quartile in the value of $p_{\Delta z}$. Most of 
the outliers are at low values of $p_{\Delta z}$, what allows to 
eliminate them by setting a suitable threshold on 
$p_{\Delta z}$ (see text and fig. \ref{90})
b) The right plot shows that it is not possible to do something 
similar using ML redshifts and $\chi^2$ as an estimator. The 
value of $\chi^2$ of the best ML fit is plotted against the error in 
the ML redshift estimation $|z_t-z_{ML}|$. The dotted line shows 
the $25\%$ high quartile in the values of $\chi^2$. One would expect 
that low values of $\chi^2$ (and therefore better fits) would correspond 
to more reliable redshifts, but this obviously is not the case. This is not 
surprising: the outliers in this figure are all due to color/reshifts 
degeneracies as the one displayed in fig. \ref{colors}, which may give an 
extremely good fit to the colors $C$, but a totally wrong redshift.}
\label{odds}
\end{figure}

Is it possible to define some `reliability estimator', similar 
to $p_{\Delta z}$ within the ML framework? The obvious choice seems to  
be $\chi^2$. 
Fig. \ref{odds}b plots the value of $\chi^2$ vs. the ML redshift error for 
the mock catalog. It is clear that $\chi^2$ is almost useless to 
pick out the outliers. The dashed line marks the upper $25\%$ quartile in 
$\chi^2$; most of the outliers are below it, at smaller $\chi^2$ 
values. In stark contrast, fig. \ref{odds}a 
plots the errors in the BPZ redshifts {\it vs.} the values of $p_{\Delta z}$. 
The lower $25\%$ quartile, under the dashed line, contains practically 
all the outliers. By setting an appropriate threshold one can virtually 
eliminate the `catastrophic errors'.

\begin{figure}[h]
\epsscale{0.5}
\plotone{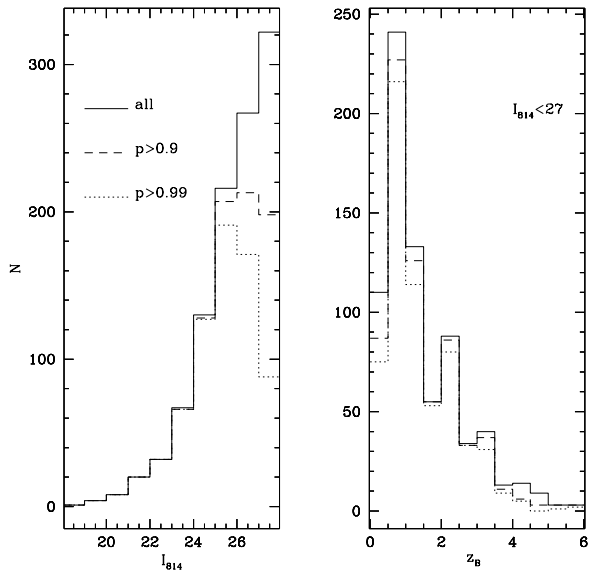}
\caption{a) On the left, histograms showing the number of galaxies over 
$p_{\Delta z}$ thresholds of $0.90$ and $0.99$ as a function of 
magnitude. It can be seen that the reliability of the photometric 
redshift estimation quickly degrades with the magnitude. 
b) The same as a) but as a function of redshift.}
\label{oddsmz}
\end{figure}

Fig. \ref{oddsmz} shows the numbers of galaxies above a given $p_{\Delta z}$
threshold in the HDF as a function of magnitude and redshifts. 
It shows how risky it is to estimate photometric redshifts using ML 
for faint, $I\gtrsim 27$ objects; the fraction of objects with possible 
catastrophic errors grows steadily with magnitude. 

  There is one caveat regarding the use of $p_{\Delta z}$ or 
similar quantities as a reliability estimator. They provide a 
safety check against the color/redshift degeneracies, since basically they 
tell us if there are other probability peaks comparable to the highest 
one, but they cannot protect us from template 
incompleteness. If the template library does not contain any spectra 
similar to the one corresponding to the galaxy, there is no indicator 
able to warn us about the unreliability of the prediction. Because of this, 
no matter how sophisticated the statistical methods become, it is 
fundamental to have a good template set, which contains---even if only 
approximately---all the possible galaxy types present in the sample. 

\begin{figure}[h]
\epsscale{0.5}
\plotone{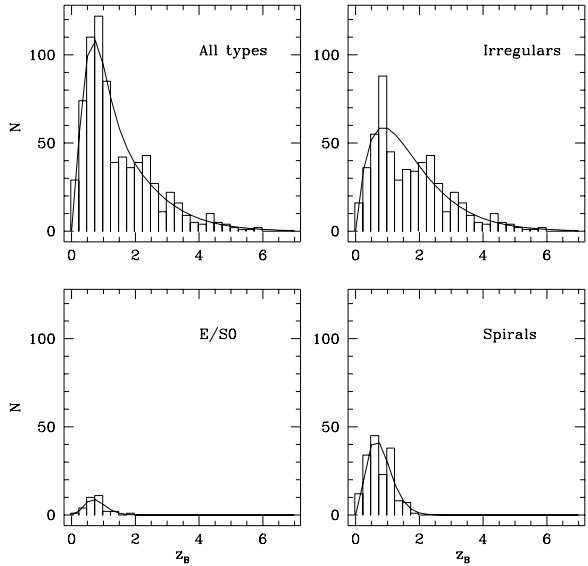}
\caption{The $z_B$ redshift distributions for the $I<27$ HDF 
galaxies divided by spectral types. The solid lines represent 
the corresponding $p(z,T)$ distributions estimated using the prior 
calibration method described in the text.}
\label{nz}
\end{figure}

Finally, fig. \ref{nz} shows the redshift distributions for the HDF 
galaxies with $I<27$. No objects have been removed on the basis of 
$p_{\Delta z}$, so the values of the histogram bins should be taken 
with care. The overplotted continous curves are the distributions 
used as priors and which simultaneously are the Bayesian fits to the 
final redshift distributions. The results obtained from the HDF will 
be analyzed in more detail, using a revised photometry, in a forthcoming 
paper.

\section{Applications}\label{appli}

As we argue above, the use of BPZ for photometric redshift 
estimation offers obvious advantages over standard ML techniques. 
However, quite often obtaining photometric redshifts is not an end 
in itself, but an intermediate step towards measuring other 
quantities, like the evolution of the star formation rate (\cite{con97}),
the galaxy--galaxy correlation function (\cite{con98},\cite{mir}), 
galaxy or cluster mass distributions (\cite{hud}), etc. The usual 
procedure consists in obtaining photometric redshifts for all the galaxies 
in the sample, using ML or the training set method, and then work with 
them as if these estimates were accurate, reliable spectroscopic redshifts. 
The results of the previous sections alert us to the dangers 
inherent in that approach, as it hardly takes into account the 
uncertainties involved in photometric redshift estimation.
In contrast, within the Bayesian framework there is no need 
to work with the discrete, point--like `best' redshift estimates. 
The whole redshift probability distribution can be taken into account, 
so that the uncertainties in the redshift estimation are accounted 
for in the final result. To illustrate this point, let's outline 
how BPZ can be applied to several problems which use photometric redshift 
estimation. 

\subsection{Spectral properties of a galaxy population}

If, instead of working with a discrete set of templates, one uses a 
spectral library whose templates depend of parameters as the 
metallicity, the star-formation history, initial mass function, etc., 
represented by $A$ in Sec \ref{bpz}, it is obvious from equation 
(\ref{cont})  that the same technique used to estimate the redshift 
can be applied to estimate any of the parameters $A$ which characterize 
the galaxy spectrum. For 
instance, let's suppose that one want to estimate the parameter 
$a_i$. Then defining $A'=\{ a_j\}, j\neq i$, we have
\begin{equation}
p(a_i|C,m_0)=\int dz\int dA' p(a_i,z,A'|C,m_0)
\propto
\int dz\int dA' p(a_i,z,A'|m_0)p(C|z,A)
\end{equation}
That is, the likelihoods $p(C|z,A)$ and the weights 
$p(a_i,z,A')\equiv p(z,A)$ are the same ones used for 
the redshift estimation (eq. \ref{cont}), only that now the 
integration is performed over the variables $z$ and $A'$ instead of 
$A$. In this way, depending 
on the template library which is being used, one can estimate galaxy 
characteristics as the metallicity, dust content, etc. 
An important advantage of this method over ML is that the estimates 
of the parameter $a_i$ automatically include the uncertainty of the 
redshift estimation, which is reflected in the final value of 
$p(a_i|C,m_0)$. Besides, by integrating the probability over all 
the parameters $A'$, one precisely  
includes the uncertainties caused by possible parameter degeneracies 
in the final result for $a_i$. It should also be noted that as many 
of the results obtained in this paper, this method can be almost 
straightforwardly applied to spectroscopical observations; one has 
only to modify the likelihood expression which compares the observed 
fluxes with the spectral template. The rest of the formalism remains 
practically identical. 

\subsection{Galaxy Clusters: member identification}

One frequent application of photometric redshift techniques is 
the study of galaxy cluster fields. The goals may be the selection of 
cluster galaxies to characterize their properties, especially at high 
redshifts, or the identification of distant, background galaxies 
to be used in gravitational lensing analysis (\cite{ben}). BPZ 
offers an effective way of dealing with such problems.

To simplify the problem, the effects of gravitational lensing on 
the background galaxies (magnification, number counts depletion, etc.) 
will be neglected (see however the next subsection). Let's suppose 
that we already have an estimate of the projected surface density 
of cluster galaxies (which can roughly be obtained without any 
photometric redshifts, just from the number counts surface 
distribution) $n_c(m_0,\vec{ r})$, where $\vec{ r}$ is the 
position with respect to the cluster center. The surface density of 
`field', non--cluster galaxies is represented by $n_f(m_0)$. For 
each galaxy in the sample we know its magnitude and colors $m_0,C$ and 
also its position $\vec{ r}$, which is now a relevant parameter in 
the redshift estimation. Following eq. (\ref{bas}) we can write
\begin{equation}
p(z|C,m_0,\vec{ r}) \propto \sum_T p(z,T|m_0,\vec{ r})p(C|z,T)
\end{equation}
A dependence on the magnitude (e.g. for the early types 
cluster sequence) could easily be included in the likelihood 
$p(C|z,T)$ if needed. The prior can be divided into the sum of two 
different terms:
\begin{equation}
p(z,T|m_0,\vec{ r})=p_c(z,T|m_0,\vec{ r})+p_f(z,T|m_0,\vec{ r})
\end{equation}
where $p_c$ represents the prior probability of the galaxy belonging 
to the cluster, whereas $p_f$ corresponds to the prior probability of 
the galaxy belonging to the general field population. The expression 
for $p_c$ can be written as 
\begin{equation}
p_c(z,T|m_0,\vec{ r})={n_c(m_0,\vec{ r}) \over n_c(m_0,\vec{ r}) + n_f(m_0)} 
p_c(T|m_0) g(z_c,\sigma z_c)
\end{equation}

The probability $p_c(T|m_0)$ corresponds to the expected galaxy mix 
fraction in the cluster, which in general will depend on the magnitude 
and will be different from that of field galaxies. The function 
$g(z_c,\sigma z_c)$ is the redshift profile of the cluster; a good 
approximation could be a gaussian with a width corresponding to the 
cluster velocity dispersion.  

The second prior takes the form
\begin{equation}
p_f(z,T|m_0,\vec{ r})={n_f(m_0) \over n_c(m_0,\vec{ r}) + n_f(m_0)}
p_f(T|m_0)p_f(z|T,m_0)
\end{equation}

which uses the priors for the general field galaxy population 
(Sec \ref{test}). Finally, the hypothesis that the galaxy belongs to 
the cluster or not can be decided about with the help of a properly 
defined $p_{\Delta z}$, or with the odds introduced in Sec \ref{bpz}.

\subsubsection{Cluster detection}

We have assumed above that the cluster redshift and its galaxy surface 
density distribution are known. However, in some cases, there is 
a reasonable suspicion about the presence of a cluster at a certain 
redshift, but not total certainty, and our goal is to confirm its existence. 
An example using ML photometric redshift estimation is shown in 
\cite{pel}. An extreme case with minimal prior information occur in 
optical cluster surveys, galaxy catalogs covering large areas of the sky 
are searched for clusters. In those cases there are no previous guesses 
about the position or redshift of the cluster, and a `blind', automatized 
search algorithm has to be used (\cite{pos}).

The prior expression used in the previous subsection offers a way to 
build such a searching method. Instead of assuming that the 
cluster redshift and its surface distribution are known, the redshift can be 
left as a free parameter $z_c$ and the expression characterizing the 
cluster galaxy surface density distribution $n_c(m_0,\vec{ r})$ can be 
parametrized using the quantities $\lambda_c$. For simplicity, 
let's suppose that 
\begin{equation}
n_c(m_0,\vec{ r})=A_c \phi(m_0,z_c) f(\vec{ r}_c, \sigma r_c)
\end{equation}
where $A_c$ is the cluster `amplitude', $\phi(m_0,z_c)$ is the 
number counts distribution expected for the cluster (which in 
general will depend on the redshift $z_c$) and $f(\vec{r}_c,\sigma r_c)$ 
represents the cluster profile, centered on $\vec{r}_c$ and with a scale 
width of $\sigma r_c$. This expression, except for the dependence 
on the redshift is very similar to that used by \cite{pos} to define 
their likelihood. Then for a multicolor galaxy sample with data 
${\bf m_0}$, ${\bf  C}$ and ${\bf \vec{ r}}$, the probability 
\begin{equation}
p(A_c,\vec{r}_c,\sigma r_c,z_c,\sigma z_c|{\bf  m_0},{\bf  C},{\bf \vec{ r}})
\end{equation}
can be developed analogously to how it was done in Sec. \ref{bpz}.
The probability assigned to the existence of a cluster at a certain 
redshift and position may be simply defined as $p(A_c|z_c,r_c)>0$. 
 
\subsection{Cluster lensing}

  It seems that the most obvious application of BPZ to cluster lensing 
analysis is the selection of background galaxies with the technique 
described in the previous subsection in order to apply the standard 
cluster mass reconstruction techniques (\cite{ks}, \cite{bro}, 
\cite{sei}, \cite{tay}). However, using Bayesian probability it 
is possible to develop an unified approach which simultaneously 
considers the lensing and photometric information in an optimal way. 

  In a simplified fashion, the problem of determining the mass 
distribution of a galaxy cluster from observables can be stated as 
finding the probability 
\begin{equation}
p(\lambda_M,\lambda_C|{\bf  e},{\bf \vec{ r}},{\bf  m_0}, {\bf  C},\lambda_G)
\label{38}
\end{equation}
where $\lambda_M$ represent the parameters which describe the cluster 
mass distribution; their number may range from a few, if the 
cluster is described with a simplified analytical model or as 
many as wanted if the mass distribution is characterized by e.g. 
Fourier coefficients (\cite{sk} ).  
$\lambda_C$ represents the cosmological parameters, which 
sensitively affect the lensing strength. The parameter set $\lambda_G$ 
represents the properties of the background galaxy population which 
affect the lensing, as its redshift distribution, number counts slope, 
etc. and it is assumed to be known previously. 
The data ${\bf e}$ correspond to the galaxy ellipticities, ${\bf\vec{ r}}$ 
to their angular positions. As above, ${\bf  m_0}, {\bf C}$ correspond 
to their colors and magnitudes. For simplicity, it will be assumed 
that the cluster and foreground galaxies have been already removed 
and we are dealing only with the background galaxy population. 

Analogous to eq. (\ref{23}), we can develop eq. (\ref{38}) as
\begin{equation}
p(\lambda_M, \lambda_C|{\bf e},{\bf \vec{ r}},{\bf m_0}, {\bf C},\lambda_G)
\propto 
p(\lambda_M) p(\lambda_C) 
\prod_i^{n_g} \int dz_i 
p(C_i,m_{0i},\vec{r_i},e_i,z_i|\lambda_M,\lambda_C,\lambda_G) 
\end{equation}
where the last factor may be written as 
\begin{equation}
p(C_i,m_{0i},\vec{r_i},e_i,z_i|\lambda_M,\lambda_C,\lambda_G)\propto 
p(e_i|z_i,\vec{r_i},\lambda_M,\lambda_C) 
p(\vec{r_i}|m_{0i},\lambda_M,\lambda_C,\lambda_G) 
p(z_i|C,m_{0i},\vec{r_i},\lambda_M,\lambda_C,\lambda_G) 
\end{equation}
The meaning of the three factors on the right side of the equation 
is the following:
$p(e_i|...)$ represents the likelihood of measuring a certain 
ellipticity $e_i$ in a galaxy given its redshift, position, etc. 
The second factor $p(\vec{r_i}|...)$ corresponds to the so called 
``Broadhurst effect'', the number counts depletion of background 
galaxies caused by the cluster magnification $\mu$ (Broadhurst 1995, 
Ben\'\i tez \& Broadhurst 1998). The last factor, $p(z_i|...)$ is 
the redshift probability, but including a correction which takes into 
account that the observed magnitude of a galaxy $m_0$ is affected by 
the magnification $\mu(\vec{r})$. It is clear that the simplified method 
outlined here is not the only way of applying Bayesian probability to 
cluster mass reconstruction. My purpose here is to show that this can 
be done considering the photometric redshifts in a integrated way with 
the rest of the information.

\subsection{Galaxy evolution and cosmological parameters}\label{gev}

As it has been shown in section (\ref{bpz}), BPZ can be used to estimate 
the parameters characterizing the joint magnitude--redshift--morphological 
type galaxy distribution. For small fields, this distribution may be 
dominated by local perturbations, and the redshift distribution may be 
`spiky', as it is observed in redshift surveys of small fields. 
However, if one were to average over a large number of fields, 
the resulting distribution would contain important 
information about galaxy evolution and the fundamental cosmological 
parameters. \cite{san} included galaxy counts as one of the four
fundamental tests of observational cosmology, although noting that the 
number-redshift distribution is in fact more sensitive to the value of 
$\Omega_0$. As \cite{gar} also notes, the color distribution of the 
galaxies in a survey hold also much more information about the process of 
galaxy evolution that the raw number counts. However, quite often the only 
method of analyzing multicolor observations is just comparing them 
with the number counts model predictions, or at most, with color 
distributions. There are several attempts at using photometric redshifts to 
study global galaxy evolution parameters (e.g. \cite{saw}, 
\cite{con97}), but so far there is not an integrated statistical 
method which would simultaneously considers all the information, magnitudes 
and colors, contained in the data, and set it against the model predictions. 

It is then clear that eq. (\ref{23}) can be used to estimate these 
parameters from large enough samples of multicolor data. If it is assumed 
that all the galaxies belong to a few morphological types, the 
joint redshift-magnitude-`type' distribution can be written as 
\begin{equation}
n(z,m_0,T)\propto {dV(z) \over dz}\phi_T(m_0)
\label{5}
\end{equation}
where $V(z)$ is the comoving volume as a function of redshift, which 
depends on the cosmological parameters $\Omega_0,\Lambda_0$ and $H_0$, 
and $\phi_T$ is the Schecter luminosity function for each morphological type
 $T$, where the absolute magnitude $M_0$ has been substituted by the 
apparent magnitude $m_0$ (a transformation which depends on the redshifts, 
cosmological parameters and morphological type). Schecter's function 
also depend on $M^{*}$, $\alpha$ and 
$\phi^{*}$, and on the evolutionary parameters 
$\epsilon$, such as the merging rate, the luminosity evolution, etc. 
Therefore, the prior probability of $z$,$A$ and $m_{0}$ depends on the 
parameters $\lambda_C=\{\Omega_0,\Lambda_0,H_0\}$, 
$\lambda_*=\{M^{*},\phi^{*}\alpha\}$ and $\epsilon$. 
As an example, let's suppose that one wants to estimate $\epsilon$, 
independently of the rest of the parameters, given the data 
${\bf D}\equiv\{D_i\}\equiv\{C_i,m_{0i}\}$. Then
\begin{equation}
p(\epsilon|{\bf D})=\int d\lambda_C d\lambda_* 
p(\epsilon,\lambda_C,\lambda_*|{\bf D})
\end{equation}
\begin{equation}
p(\epsilon|{\bf D})\propto 
\int d\lambda_C d\lambda_* 
p(\epsilon,\lambda_C,\lambda_*)
\prod_i\int dz_i\sum_{T} p(z_i,T,m_{0i}|\epsilon,\lambda_C,\lambda_*)
p(C_i|z_i,T)
\label{6}
\end{equation} 
The prior $p(z_i,T,m_{0i}|\epsilon,\lambda_C,\lambda_*)$ can be derived 
from $n(z,m_0,T)$ in eq. (\ref{5}). The 
prior $p(\epsilon,\lambda_C,\lambda_*)$ 
allows to include the uncertainties derived from previous observations or 
theory in the values of these parameters, even when they are strongly 
correlated among themselves, as in the case of the Schecter function 
parameters $\lambda_*$. The narrower the prior 
$p(\epsilon,\lambda_C,\lambda_*)$ is, the less `diluted' the 
probability of $\epsilon$ and the more accurate the estimation. 

\section{Conclusions}\label{conc}

Despite the remarkable progress of faint galaxy spectroscopical 
surveys, photometric redshift techniques will become increasingly 
important in the future. The most frequent approaches, the 
template--fitting and empirical training set methods, present several 
problems related which hinder their practical application. Here it is 
shown that by consistently applying Bayesian probability to photometric 
redshift estimation, most of those problems are efficiently solved. 
The use of prior probabilities and Bayesian marginalization allows the 
inclusion of valuable information as the shape of the redshift 
distributions or the galaxy type fractions, which is usually ignored 
by other methods. It is possible to characterize the accuracy of 
the redshift estimation in a way with no equivalents in other statistical 
approaches; this property allows to select galaxy samples for 
which the redshift estimation is extremely reliable. In those cases when 
the {\it a priori} information is insufficient, it is shown how to 
`calibrate' the prior distributions, using even the data under 
consideration. In this way it is possible to determine the properties 
of individual galaxies more accurately and simultaneously estimate 
their statistical properties in an optimal fashion.

The photometric redshifts obtained for the Hubble Deep Field using 
optical and near-IR photometry show an excellent agreement with the 
$\sim 100$ spectroscopic redshifts published up to date in the 
interval $1<z<6$, yielding a rms error $\Delta z_B/(1+z_{spec}) = 0.08$ 
and no outliers. Note that these results, obtained with an empirical 
set of templates, have not been reached by minimizing the difference 
between spectroscopic and photometric redshifts (as for 
empirical training set techniques, which may lead to an overestimation 
of their precision) and thus offer a reasonable estimate of the 
predictive capabilities of BPZ. 
The reliability of the method is also tested by estimating redshifts 
in the HDF but restricting the color information to the UBVI filters; 
the results are shown to be more reliable than those obtained with 
the existing techniques even including the near-IR information. 

The Bayesian formalism developed here can be generalized to deal 
with a wide range of problems which make use of photometric redshifts. 
Several applications are outlined, e.g. the estimation of individual 
galaxy characteristics as the metallicity, dust content, etc., or the 
study of galaxy evolution and the cosmological parameters from large 
multicolor surveys. Finally, using Bayesian probability it is possible to 
develop an integrated statistical method for cluster mass reconstruction 
which simultaneously considers the information provided by gravitational 
lensing and photometric redshift estimation. 

\acknowledgements 

I would like to thank Tom Broadhurst and Rychard Bouwens for careful reading 
the manuscript and making valuable comments. Thanks also to Alberto 
Fern\'andez-Soto and collaborators for kindly providing me with the HDF 
photometry and filter transmissions, and to Brenda Frye for help with 
the intergalactic absorption correction. The author acknowledges a 
Basque Government postdoctoral fellowship.

\newpage
\begin{deluxetable}{lrrr}
\tablewidth{33pc}
\tablecaption{Parameters of the priors $p(z|T,m_0)$(see text)}
\tablehead{
\colhead{Spectral type} & \colhead{$\alpha_t$} &
\colhead{$z_{0t}$} &\colhead{$k_{mt}$}
}
\startdata
E/S0    &  $2.26\pm 0.05$   & $0.48\pm 0.03$  & $0.061\pm 0.06$ \nl
Sbc,Scd &  $1.71\pm 0.04$   & $0.44\pm 0.02$  & $0.044\pm 0.002$\nl
Irr     &  $1.125\pm 0.015$ & $0.038\pm 0.01$ & $0.178\pm 0.002$\nl
\enddata
\end{deluxetable}


\begin{thebibliography}{}

\bibitem[Ben\'\i tez \& Broadhurst 1998]{ben} 
Ben\'\i tez, N. \& Broadhurst, T. 1998, in preparation

\bibitem[Bretthorst] {}Bretthorst, L. 1988, Bayesian Spectrum 
Analysis and Parameter Estimation, Lecture Notes Series, vol. 48, 
Springer--Verlag

\bibitem[Bretthorst] {}Bretthorst, L. 1990, Journal of 
Magnetic Resonance, 88, 552

\bibitem[Broadhurst 1995]{bro} Broadhurst, T. 1995, astro-ph/9511150

\bibitem[Brunner et al. 1997]{bru} Brunner, R. J., 
Connolly, A. J., Szalay, A. S., \& Bershady, M. A. 1997, 
\apjl, 482, L21 

\bibitem[Bruzual \& Charlot 1993]{bc} Bruzual A., G. \& 
Charlot, S. 1993, \apj, 405, 538 

\bibitem[Coleman, Wu, \& Weedman 1980]{cww} Coleman, G. D., 
Wu, C.-C., \& Weedman, D. W. 1980, \apjs, 43, 393 

\bibitem[Connolly et al. 1995]{con95} Connolly, A. J., 
Csabai, I., Szalay, A. S., Koo, D. C., Kron, R. G., \& Munn, J. A. 1995, 
\aj, 110, 2655 

\bibitem[Connolly et al. 1997]{con97} Connolly, A. J., 
Szalay, A. S., Dickinson, M., Subbarao, M. U., \& Brunner, R. J. 1997, 
\apjl, 486, L11 

\bibitem[Connolly, Szalay, \& Brunner 1998]{con98} Connolly, 
A. J., Szalay, A. S., \& Brunner, R. J. 1998, \apjl, 499, L125 

\bibitem[Dey et al. 1998]{dey} Dey, A., Spinrad, H., Stern, 
D., Graham, J. R., \& Chaffee, F. H. 1998, \apjl, 498, L93 

\bibitem[Dickinson et al. 1998]{dic}
Dickinson, M. et al. 1998, in preparation

\bibitem[Ellis 1997]{ellis} Ellis, R. S. 1997, 
\araa, 35, 389 

\bibitem[Franx et al. 1997]{fra} Franx, M., Illingworth, G. 
D., Kelson, D. D., Van Dokkum, P. G., \& Tran, K.-V. 1997, \apjl, 486, L75 

\bibitem[Frye, Broadhurst \& Ben\'\i tez 1998]{fry} Frye, B., 
Broadhurst, T. \& Ben\'\i tez, N. 1998, in preparation

\bibitem[Fern\'andez-Soto, Lanzetta \& Yahil 1998]{fsoto}
Fern\'andez-Soto, A., Lanzetta, K.M. \& Yahil, A. 1998, 
to appear in ApJ, astro-ph/9809126 

\bibitem[Gardner 1998]{gar} Gardner, J. P. 1998, 
\pasp, 110, 291 

\bibitem[Giallongo et al. 1998]{gia} Giallongo, E., 
D'Odorico, S., Fontana, A., Cristiani, S., Egami, E., Hu, E., \& McMahon, R. 
G. 1998, \aj, 115, 2169 


\bibitem[Gwyn \& Hartwick 1996]{gwy} Gwyn, S. D. J. \& 
Hartwick, F. D. A. 1996, \apjl, 468, L77 

\bibitem[Hogg et al. 1998]{hog} Hogg, D. W., et al. 1998, 
\aj, 115, 1418 

\bibitem[Hudson et al. 1998]{hud} Hudson, M. J., Gwyn, S. 
D. J., Dahle, H., \& Kaiser, N. 1998, \apj, 503, 531 

\bibitem[Jaynes 1998]{} E.T. Jaynes, ``Probability theory: The logic 
of science'', to be published in Cambridge University Press. A preliminary 
version can be obtained from Thomas Loredo's web page at {\it 
http://astrosun.tn.cornell.edu/staff/loredo/bayes/}. 

\bibitem[Kaiser \& Squires 1993]{ks} Kaiser, N. \& 
Squires, G. 1993, \apj, 404, 441 

\bibitem[Kinney et al. 1996]{kin} Kinney, A. L., Calzetti, 
D., Bohlin, R. C., McQuade, K., Storchi-Bergmann, T., \& Schmitt, H. R. 
1996, \apj, 467, 38 

\bibitem[Kodama, Bell \& Bower 1998]{kod} Kodama, T., Bell, E. F. 
\& Caldwell, N. 1998, to appear in \mnras, astro-ph/9806120

\bibitem[Koo 1985]{koo} Koo, D.C. 1985, AJ, 90, 418

\bibitem[Lanzetta, Yahil \& Fern\'andez-Soto 1996]{lan}
Lanzetta, K.M., Yahil, A.\& Fern\'andez-Soto, A. 1996, Nature, 
381, 759

\bibitem[Liu \& Green 1998]{liu} Liu, C. T. \& Green, R. F. 
1998, \aj, 116, 1074 

\bibitem[Loredo] {}Loredo, T. 1990, in {\it Maximum Entropy 
and Bayesian Methods, Darmouth}, Ed. P., Fougere, Kluwer Academic 
Publishers, p. 81-142

\bibitem[Loredo]{} Loredo, T. 1992, in {\it Statistical 
Challenges in Modern Astronomy}, ed. E.D. Feigelson and G.J. Babu 
(New York: Springer-Verlag) pp. 275--297 

\bibitem[Madau 1995]{mad} Madau, P. 1995, \apj, 441, 18 

\bibitem[Miralles \& Pell\'o 1998]{mir} 
Miralles, J.M. \& Pell\'o, R. 1998, astro-ph/9801062

\bibitem[Pell\'o et al. 1996]{pel} Pello, R., Miralles, J.M., 
Le Borgne, J.-F., Picat, J.-P., Soucail, G., \& Bruzual, G. 1996, 
\aap, 314, 73 

\bibitem[Postman et al. 1996]{pos} Postman, M., Lubin, L. 
M., Gunn, J. E., Oke, J.B., Hoessel, J. G., Schneider, D. P., \& 
Christensen, J. A. 1996, \aj, 111, 615 

\bibitem[Press et al. 1992]{nr} Press, W. H., Teukolsky, 
S. A., Vetterling, W. T., \& Flannery, B. P. 1992, Numerical recipes in 
FORTRAN. The art of scientific computing, Cambridge: University Press. 

\bibitem[Sandage 1961]{san}Sandage, A. 1961a, ApJ, 133, 355

\bibitem[Sawicki, Lin, \& Yee 1997]{saw} Sawicki, M. J., 
Lin, H., \& Yee, H. K. C. 1997, \aj, 113, 1 

\bibitem[Seitz, Schneider, \& Bartelmann 1998]{sei} Seitz, 
S., Schneider, P., \& Bartelmann, M. 1998, \aap, 337, 325 

\bibitem[Sodr\'e \& Cuevas 1997]{sod} Sodr\'e, L. \& Cuevas, H. 
1997, \mnras, 287, 137

\bibitem[Squires \& Kaiser 1996]{sk} Squires, G. \& 
Kaiser, N. 1996, \apj, 473, 65 

\bibitem[Taylor et al. 1998]{tay} Taylor, A. N., Dye, S., 
Broadhurst, T. J., Benitez, N., \& Van Kampen, E. 1998, \apj, 
501, 539 

\bibitem[Wang, Bahcall \& Turner 1998]{wan} Wang, Y., 
Bahcall, N. \& Turner, E.L. 1998, to appear in AJ, 
astro-ph/9804195

\bibitem[Williams et al. 1996]{wil} Williams, R. E., et al.
1996, \aj, 112, 1335 

\bibitem[Yee 1998]{yee} Yee, H.K.C. 1998, to appear in the proceedings of 
the Xth Rencontres de Blois, astro-ph/9809347


\end{thebibliography}
\end{document}